\documentclass[pre,twocolumn,superscriptaddress,preprintnumbers,aps,showpacs,amsmath,amssymb]{revtex4}

\usepackage[pdftex]{graphicx} 
\usepackage{dcolumn}
\usepackage{bm}
\usepackage{color}
\usepackage{longtable}
\usepackage{xr}
\usepackage{subfigure}

\newcommand{\beq}{\begin{equation}}
\newcommand{\eeq}{\end{equation}}
\newcommand{\beqa}{\begin{eqnarray}}
\newcommand{\eeqa}{\end{eqnarray}}
\newcommand{\aver}[1]{\left\langle {#1}\right\rangle}

\newcommand{\bem}{\begin{math}}
\newcommand{\eem}{\end{math}}
\newcommand{\rar}{{\rightarrow}}

\newcommand{\bfr}{{\bf r}}
\newcommand{\bfu}{{\bf u}}
\newcommand{\bfp}{{\bf p}}

\newcommand{\bff}{{\bf f}}
\newcommand{\bfv}{{\bf v}}
\newcommand{\bfe}{{\bf e}}

\newcommand{\bsigma}{{\boldsymbol \sigma}}

\newcommand{\bomega}{{\boldsymbol \omega}}

\def\strutdepth{\dp\strutbox}
\def\nw#1{\strut\vadjust{\kern-\strutdepth\vtop to0pt{\vss\hbox to\hsize
{\hskip\hsize\hskip5pt$\leftarrow$\hss\strut}}}{\em #1}}
\usepackage{multirow}
\def\tl#1{\textcolor{black}{#1}}

\def\mean#1{\left< #1 \right>}
\allowdisplaybreaks[3]

\def\NY#1{\textcolor{black}{#1}}
\def\NYY#1{\textcolor{black}{#1}}

\usepackage{amsmath}	
\begin{document}

\preprint{}

\title{
Hydrodynamic interactions in dense active suspensions:
from polar order to dynamical clusters
}

\author{Natsuhiko Yoshinaga}
\email[E-mail: ]{yoshinaga@tohoku.ac.jp}
\affiliation{
WPI - Advanced Institute for Materials Research, Tohoku University,
Sendai 980-8577, Japan}
\affiliation{
MathAM-OIL, AIST,
Sendai 980-8577, Japan}
\affiliation{
The Kavli Institute for Theoretical Physics, University of California, Santa Barbara, CA 93106, USA 
}

\author{Tanniemola B. Liverpool}
\email[E-mail: ]{t.liverpool@bristol.ac.uk}
\affiliation{
The Kavli Institute for Theoretical Physics, University of California, Santa Barbara, CA 93106, USA 
}
\affiliation{School of Mathematics, University of Bristol, Bristol, BS8 1TW, UK}
\affiliation{BrisSynBio, Tyndall Avenue, Bristol, BS8 1TQ, UK}

 \begin{abstract}
We study the role of hydrodynamic interactions in the collective behaviour of collections of microscopic active particles
suspended in a fluid. 
  We introduce a novel calculational framework that allows us to separate the different contributions to their collective dynamics from hydrodynamic interactions on different length scales. Hence we are able to systematically show that lubrication forces when the particles are very close to each other play as important a role as long-range hydrodynamic interactions in determining their many-body behaviour. 
   We find that  motility-induced phase separation is suppressed by near-field interactions, leading to open gel-like clusters rather than dense clusters. Interestingly, we  find a globally polar ordered phase appears for neutral swimmers with no force dipole that is enhanced by near field lubrication forces in which the collision process rather than long-range interaction dominates the alignment mechanism.
 \end{abstract}








\maketitle


Active materials are condensed matter systems self-driven out of equilibrium by  
components that convert stored energy into movement. They  have generated much interest  recently, both as inspiration for 
new smart materials and as a framework to understand 
aspects of  cell motility~\cite{Marchetti:2013,Toner2005,ramaswamy:2010}.  
They are characterised by interesting non-equilibrium collective phenomena, such as swirling, alignment, pattern formation, dynamic cluster formation and phase separation \cite{vicsek:1995,Henricus:2012,Cates:2012,Palacci:2013,Bricard:2013}. 
Theoretical descriptions of active systems range from continuum models~\cite{Marchetti:2013,Bertin:2013a} to discrete collections of self-propelled active particles~\cite{vicsek:1995}. 
An influential classification of self-propelled active particle systems has been to group them into 
 {\it dry} and {\it wet} systems~\cite{Marchetti:2013}.  Dry systems do not have momentum conserving dynamics (e.g, Vicsek models~\cite{vicsek:1995,Bertin:2013a} and Active Brownian particle (ABP) models interacting via soft {\em repulsive} potentials~{\cite{Fily:2012,Buttinoni:2013,Redner:2013}}), while wet systems conserve momentum via a coupling to a fluid (e.g. Squirmers  driven by surface deformations~\cite{Lighthill:1952,blake:1971,Pak:2014} and Janus particles driven by surface chemical reactions~\cite{golestanian:2005}) leading to hydrodynamic interactions between active particles.
Dealing with hydrodynamics leads to significant technical hurdles;  as the motion of a self-propelled swimmer is affected by other particles due to both fluid flow and pressure, and even
the two-body interaction between spherical squirmers in close proximity (near-field) is non-trivial, requiring sophisticated numerical analyses~\cite{ishikawa:2006,Llopis:2006,Ishikawa:2008a, Ishimoto:2013,Li:2014,SharifiMood:2015,Papavassiliou:2016}. 
Therefore,  converting this into an understanding of collective
behaviour remains a significant challenge~\cite{Mucha:2004}.
Because numerical simulations with hydrodynamics require significantly more computational power, studies of these systems have
relatively few particles or low resolution of fluid flow~\cite{Molina:2013,Zoettl:2014,Matas-Navarro:2014,Delfau:2016}.
Hence, far-field approximations (swimmers as point multipoles
) 
~\cite{Spagnolie:2012} are often used to account for
hydrodynamic interactions~\cite{Saintillan:2015}.
This clearly breaks down when the swimmers are close to one another, limiting the validity of such studies to very dilute suspensions.

The appearance of dynamical clusters \cite{Theurkauff:2012,Theurkauff:2012,Buttinoni:2013} in recent experiments on active particles has generated much interest. 
This has been linked to a clustered state is observed in two-dimensional (2D) ABP systems called motility-induced phase
separation (MIPS)~\cite{Fily:2012,Buttinoni:2013,Redner:2013,peruani:2006} and for squirmers confined between walls \cite{Zoettl:2014,Blaschke:2016}. A major difference however is finite size clusters in experiments~\cite{Theurkauff:2012,Palacci:2013,Buttinoni:2013} compared to the infinite cluster formed in MIPS.
In addition, recent simulations have shown that  clusters are absent in 
2D squirmer suspensions and in a 2D squirmer monolayer embedded in a 3D fluid~\cite{Matas-Navarro:2014,Saffman:1975}.
\tl{
While attractive interactions can lead to 
clustering  \cite{Saha:2013,Navarro:2015,Alarcon:2016}, here 
we study swimmers with purely repulsive interactions to see role of hydrodynamics in active 
cluster formation.
}

\begin{figure}[hb]
\begin{center}
\includegraphics[width=0.40\textwidth]{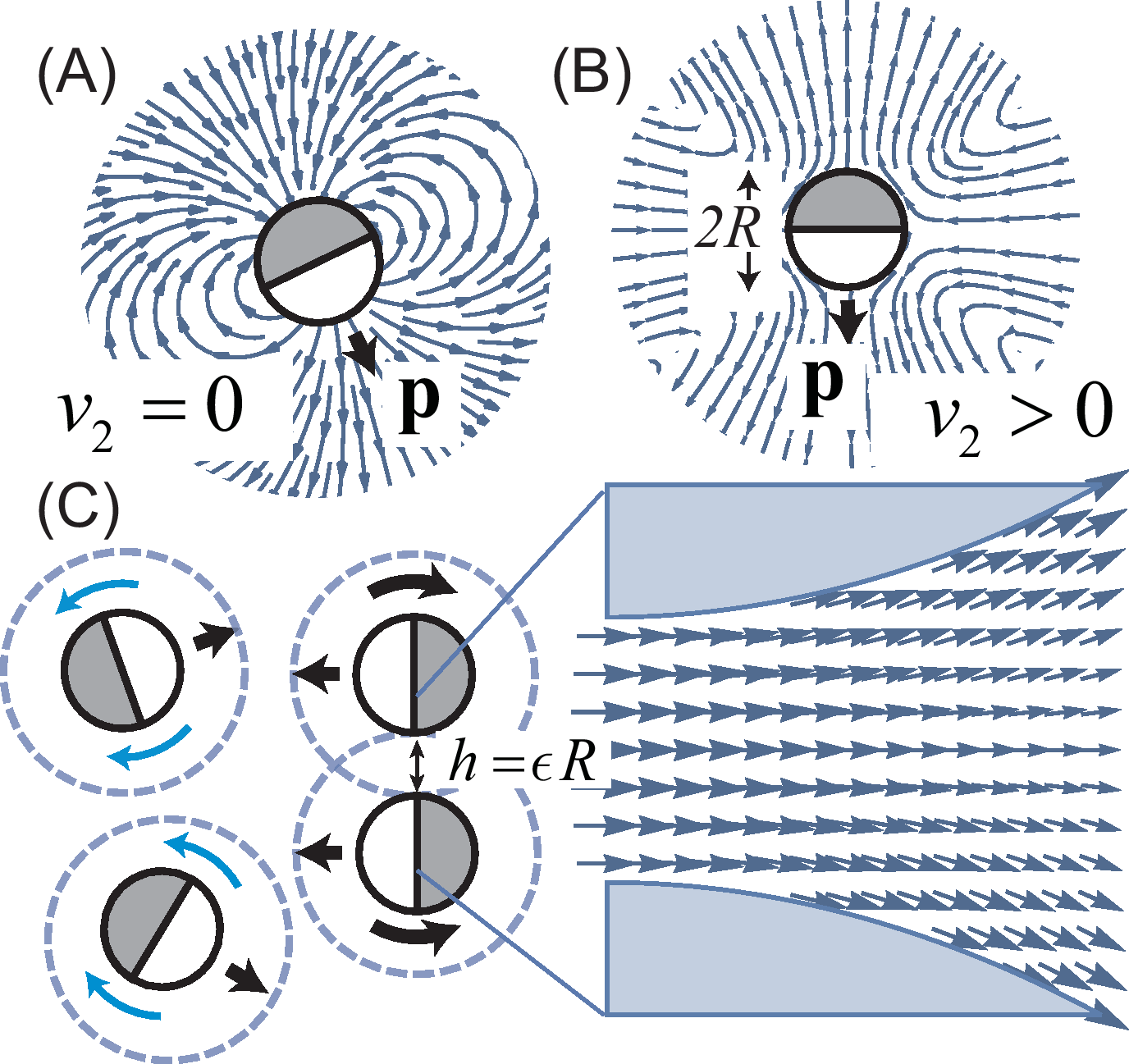}
 \caption{
 (Colour Online) Schematic interaction between swimmers.
 Each particle creates a leading order (A) quadrupolar or (B) dipolar flow.
 (C) When two particles are very close to each other, lubrication flow dominates the interactions.
\label{fig.schematics}
}
\end{center}
\end{figure}

It is accepted from continuum arguments that the polar state is generically unstable for wet active systems~\cite{Marchetti:2013,Toner2005,ramaswamy:2010}, however recent simulations of wet active particles have  raised  the interesting possibility of other 
continuum limits in these systems.
A polar state has been observed for neutral squirmers with no force dipole  with 3D hydrodynamics, but 2D  motion~\cite{Ishikawa:2008a} and in 3D~\cite{Evans:2011,Alarcon:2013,Oyama:2016}.
\tl{It has been suggested  \cite{Ishikawa:2008a} that near-field effects {\it
enhance} the polar state although there are hints that 
 far-field effects also play a role~\cite{Ishikawa:2008a}}.
These results are limited by  \tl{relatively} few particles so it is natural to ask if the polar state is present in the thermodynamic limit. 

\begin{figure}[htbp]
\begin{center}
\includegraphics[width=0.5\textwidth]{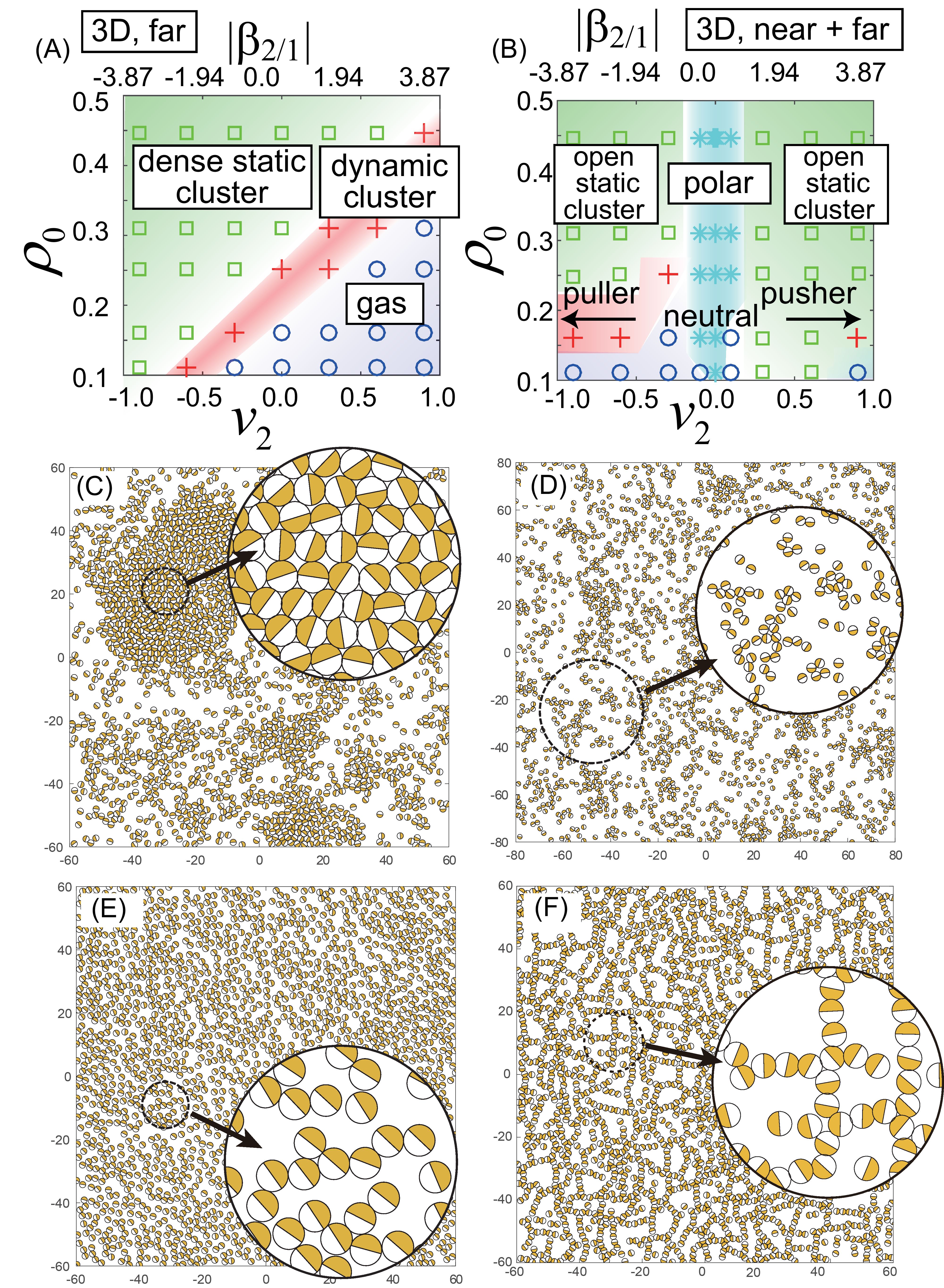}
 \caption{
 (Colour Online)
 The state diagram of squirmers with density ($\rho_0$) and dipolar strength ($v_2$) with (A) far field
 hydrodynamic interactions and (B) both far field and near field with $N=2048$ particles.
 Snap shots of (A) the static dense cluster state for a far-field-only system
 ($\phi=0.447$, $v_2=0$), (B) the dynamic cluster state($\phi=0.251$, $v_2=0$), (C) the
 polar state ($\phi=0.447$, $v_2=0$), and (D) the static open cluster state for a near+far system($\phi=0.447$, $v_2=0.9$).
\label{fig.Phase.Diagram}
}
\end{center}
\end{figure}

In this letter, we systematically construct equations of
motion for wet active particles, namely, the dynamics of their position and
orientation.  
One of our aims is to provide a computationally tractable model 
of comparable complexity to ABP  which takes account of hydrodynamic interactions 
of particles in close proximity~\cite{ishikawa:2006,Swan:2011}. 
Using it, we study a suspension of \tl{force/torque free} repulsive  {\em
spherical} squirmers and obtain  the phase behaviour summarized in
Fig. \ref{fig.Phase.Diagram} as a function of density $\rho_0$.
In studying the phase behaviour, we have emphasised the dependence on the sign of the force dipole  ($v_2$) and contrasted them to neutral swimmers with force quadrupole and no force dipole ($v_2=0$).  
 We find significant differences between the hydrodynamic interactions with
 and without near-field effects.
\tl{The phase behaviour of neutral swimmers ($v_2=0$)
with only {\em far-field} interactions are similar to those of
 ABPs since there are no collision-induced reorientations. 
 Upon including near field effects, we obtain phase diagrams characterised by 
at low densities,  a disordered  `gas'  state and at  higher
densities, the emergence of stable `static clusters' except for neutral
swimmers~\footnote{See ~\cite{SM} for the definitions of the
 polar order  and the clusters} which spontaneously develop polar order.
 Dense static clusters,
 present in far-field only system,
  are suppressed by the near-field
 interactions, leading to open gel-like clusters.
 In between the gas and static cluster are phases of `dynamic clusters'  of finite size that
 exchange particles with bulk.
  While the boundaries  between different clustered phases are qualitative and threshold-dependent, the 
 boundaries between the polar state and other states has all the features associated with a phase transition.
 }

Each particle (squirmer) is characterized by its position and orientation $({\bf
r}^{(i)} , {\bf p}^{(i)} )$ with dynamics given by
 \begin{align}
\dot{{\bf r}}^{(i)} 
&=
{\bf u}^{(i)} 
; \;\;
\dot{{\bf p}}^{(i)} 
=
{\bm \omega}^{(i)} 
\times {\bf p}^{(i)} ; \; \; | {\bf p}^{(i)} | = 1 \; , 
  \label{janus.dotr.u}
\end{align}
The translational and angular velocities of each particle, ${\bf u}^{(i)} $ and
${\bm \omega}^{(i)} $ respectively,  are obtained by solving 
 for
 \NY{
the fluid mediated interaction between all pairs of particles 
 }
The fluid is taken as incompressible with vanishing Re :
\begin{align}
 \eta \nabla ^2 {\bf v} - \nabla p 
 &=
 0 \; ; \, 
 \; \;
 \nabla \cdot {\bf v}
 =
 0
 \label{eq:stokes1}
\end{align}
where $\eta$ is viscosity, ${\bf v}({\bf r})$ is the
velocity, and $p ({\bf r})$ the pressure. The boundary
condition on the swimmer surface is a sum of rigid translational, ${\bf u}$ and rotational, ${\bm \omega}$
motion and an active slip flow, ${\bf v}_s$ driving self-propulsion:
\begin{align}\left. {\bf v} \right|_{{\bf r} = {\bf R}} &=
{\bf u} + {\bm \omega} \times {\bf R} + {\bf v}_s 
\label{BC} \\
  {\bf v}_s &=
 \sum_{l =1}^{\infty}
 \sum_{m=-l}^{l}
 \left[ v_{lm} {\bf \Psi}_{lm} (\theta,\varphi) + w_{lm} {\bf \Phi}_{lm} (\theta,\varphi)
\right],
\label{janus.intro.slip}
\end{align}
for a swimmer with centre at the origin.
The fluid velocity vanishes
 at infinity, ${\bf
v}|_{r \rightarrow \infty} = 0$,
with $\theta$ the polar angle with the $z$-axis and azimuthal, $\varphi$ with the $x$-axis on the $xy$-plane.
The slip velocity ${\bf v}_s$ can be very efficiently expanded in the 
tangential vector spherical harmonics, ${\bf \Psi}_{lm}$ and ${\bf \Phi}_{lm}$~\cite{Hill:1954,SM}.
The second term in (\ref{janus.intro.slip}) represents rotational slip
\NY{around the swimmer axis}
associated with spinning motion which we neglect in the following 
and from now on set $w_{lm}=0$.
The swimmer axis
$
 {\bf p}
$
 is a unit vector (see Fig.~\ref{fig.schematics}).
For uniaxial particles, $v_{lm}$ is a function of a magnitude $v_l$
and the swimmer orientation ${\bf p}$~\cite{SM}. 
\tl{An isolated squirmer moves with the velocity
$
 {\bf u}_0^{(i)} = u_0 {\bf p}^{(i)}
 $ with $
 u_0 =
 - \frac{2}{3} \sqrt{\frac{3}{4\pi}} v_1
 $.}


\tl{Given two squirmers, separated by ${\bf r}_{ij}$, the flow field generated by one will
affect the other and hence lead to modification of the self-propulsion
velocities. We split the problem into two parts, a force and torque,   acting
on the sphere with: 1st, slip boundary conditions without
translational and rotational motion (${\bf F}^{(p)},{\bf T}^{(p)}$, the passive problem), and 2nd with the non-slip boundary
conditions undergoing rigid-body motion ${\bf u}^{(i)}$ and ${\bm
\omega}^{(i)}$ (${\bf F}^{(a)},{\bf T}^{(a)}$, the active problem)~\cite{SM}.
The force and torque-free conditions imply, 
$
 {\bf F}^{(a)} + {\bf F}^{(p)} =0
 $ and $
 {\bf T}^{(a)} + {\bf T}^{(p)} =0
 $ which determine ${\bf u}^{(i)}$ and ${\bm
\omega}^{(i)}$.
The problems can be solved exactly 
for pairs of particles in two asymptotic limits : (1)
when their separation, $h_{ij}=r_{ij}-2R, r_{ij}=|{\bf r}_{ij}|$ is much less than their radius (near-field) and  (2) when their separation is much greater than their radius (far-field).
For arbitrary separations
between particles, we interpolate between the two limits; far-field
and near field,  using the $\tanh$ function.
There is long history of calculation of the passive problem~
\cite{Jeffrey:1984,Kim:1991}. 
Here we  compute for the first time
the active problem for both far-field and
near-field  in the general setting.
Previous near-field active results 
have been obtained only for axisymmetric surface flow-fields \NY{\cite{ishikawa:2006}}.
It should be noted that to obtain the velocity and angular
velocity for collections of swimmers, one must solve 
the active/passive problems for  {\it all} possible relative orientations 
which has not been achieved before \NY{\cite{Papavassiliou:2016}}.
}

\tl{
For a pair of
squirmers (labelled $i,j$) with arbitrary positions (and orientation),
we define a spherical coordinate system : relative separations
$r_{ij}=|{\bf r}_{ij}|$, polar angles $\theta_{ij}$ and
azimuthal angles $\varphi_{ij}$. Using it, a general form for the velocities valid in both 
far and near field limits is
\begin{align}
{\bf u}^{(i)}
 &=
 {\bf u}_0^{(i)} \lambda^{(i)} +
 \sum_{j \neq i, l,m }
 \left[
 u_{lm,\parallel}^{(ji)}
 {\bf Y}_{lm}^{(ji)} 
 +
u_{lm,\perp}^{(ji)}
 {\bf \Psi}_{lm}^{(ji)} 
 \right]
 \label{trans.vel}
 \\
{\bm \omega}^{(i)} 
&=
 \sum_{j \neq i, l,m}
\omega_{lm}^{(ji)}
 {\bm \Phi}_{lm}^{(ji)}
 .
 \label{ang.vel}
\end{align}
where  ${\bf
 \Psi}_{lm}^{(ji)}={\bf \Psi}_{lm} ( \theta_{ji}, \varphi_{ji})$, ${\bf
 \Phi}_{lm}^{(ji)}={\bf \Phi}_{lm} ( \theta_{ji}, \varphi_{ji})$, ${\bf
 Y}_{lm}^{(ji)}={\bf Y}_{lm} ( \theta_{ji}, \varphi_{ji})$ with the
 ${\bf
Y}_{lm}(\theta,\varphi)$ the normal vector spherical harmonics.
For the far-field, $u_{1,m,\parallel}^{(ji)} \sim u_{1,m,\perp}^{(ji)} \sim
r_{ij}^{-3}$ (quadrupole), and  $u_{2,m,\parallel}^{(ji)} \sim r_{ij}^{-2}$, and $\omega_{2,m}^{(ji)}
\sim r_{ij}^{-3}$ (dipole).
For the near-field, $u_{lm,\parallel}^{(ji)} \sim
\epsilon \log \epsilon$ and $u_{lm,\perp}^{(ji)} \sim \omega_{lm}^{(ji)} \sim \epsilon^0$ with $\epsilon=(r_{ij}-2R)/r_{ij}$~\cite{SM}.
$\lambda^{(i)}=1$ when the $i$th particle is away from near-field region
of any other particles and $\lambda^{(i)}=0$ otherwise. 
}

\tl{
Equations
(\ref{janus.dotr.u}), (\ref{ang.vel}), and
(\ref{trans.vel})  form a closed complete dynamical system.
Using them, we performed numerical simulations of $N$ identical particles of radius
$R$ 
with periodic boundary conditions.
Figure \ref{fig.Phase.Diagram} shows various state points of this model as a
function of the density, $\rho_0= \pi R^2 N/L^2 $
and the force dipole strength $v_2$.
Most have $N=2048$ unless specified otherwise.
Defining, the average distance between two particles, $\xi = R \sqrt{\pi/\rho_0}$,
we vary $\xi$ from $\xi \simeq 2.65$ to $\xi \simeq 5.30$.
We set $v_1 =1$ for all swimmers and thus $u_0 \simeq 0.32$.}

\tl{
The size of a particle is chosen to be of unit length, thus we set $R=1$.
The time scale is normalised by the time for an isolated squirmer to
move a half of its body length, that is, $\tau_0 = R/u_0$.
There is a time scale associated with  collisions, $\tau_m = \xi/u_0$.
 We vary the time scale from $\tau_m \simeq 8$ to $\tau_m \simeq 17$.
}
We consider motion restricted to 2D {\em but} interacting via 3D hydrodynamic interactions.
We neglect the modes  with $l \ge 3$.
We note that for 3D hydrodynamics projected onto 2D, pushers and pullers are not identical; 
the interaction at the front and the back is stronger than that at the side.
As a result, pullers, on average attract nearby objects.
\tl{We find global phase
separation of active particles with repulsive interactions, i.e. MIPS,
is suppressed by near-field hydrodynamics and we find instead
 networks of open clusters for a large range of intermediate
densities. We see a gel-like extended state at high enough densities.}

\begin{figure}[h]
\begin{center}
\includegraphics[width=0.50\textwidth]{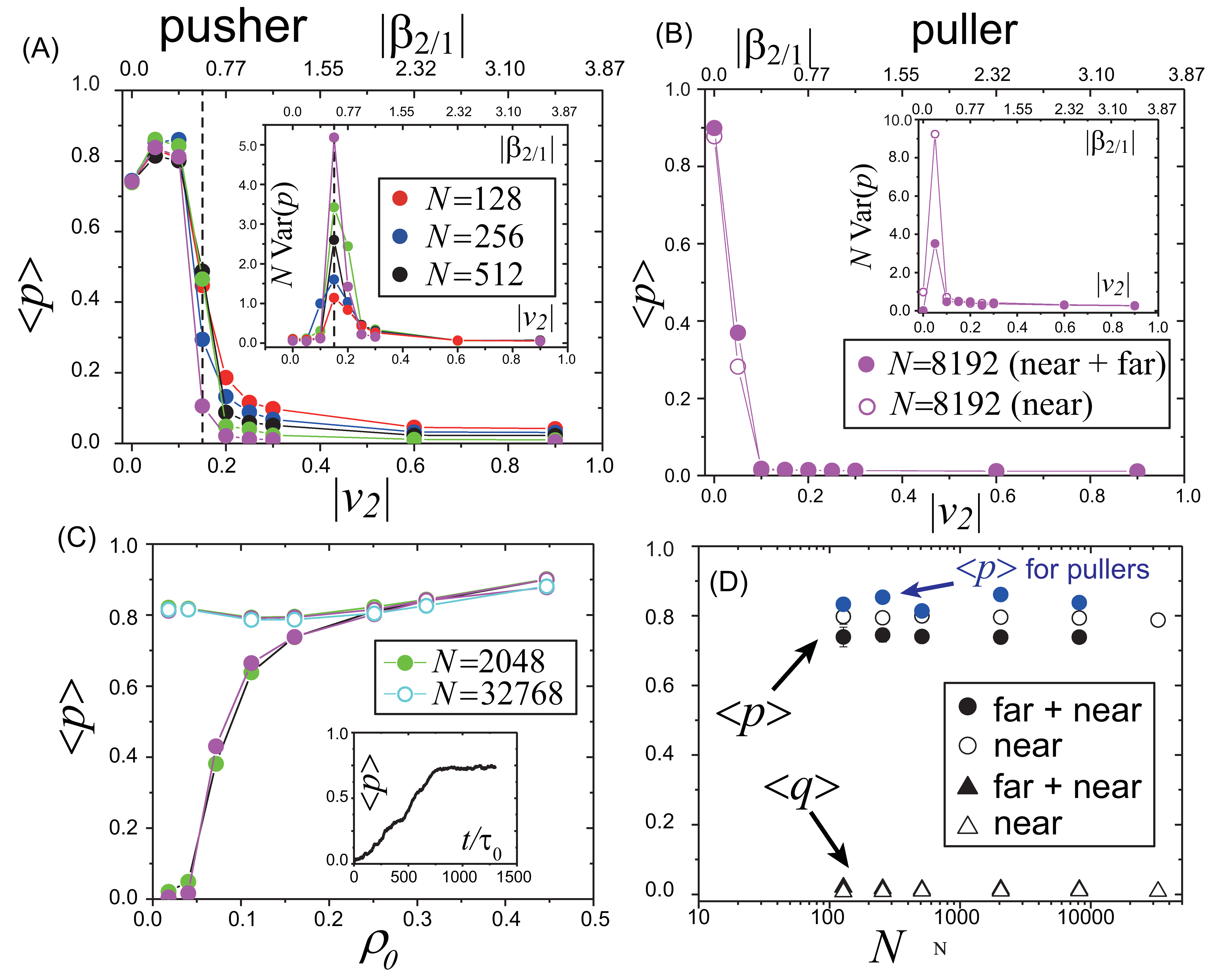}
 \caption{
 (Colour Online)
 (A, B) The mean polarity as a function of the dipolar slip
 flow $v_2$ on a squirmer at $\rho_0=0.16$ ($\xi = 4.43$).
 The corresponding values of the squirmer index $\beta_{2/1}$ is shown
 at the upper axis.
 The polar order $\mean{p}$ for pushers (A) and pullers (B).
 The insets in (A) and (B) show the variance of the polar order.
 Closed circles indicate the system with the near and far fields whereas
 open circles indicate the near-field-only system.
 (C) the mean polarity, $\mean{p}$ for neutral swimmers
 $v_2=0$ vs density, $\rho_0$ (inset: time evolution of $\mean{p}$).
The solid (open) symbols correspond to the simulations with (without)
 the far-field interaction.
 Both include the near field interaction.
(D) The system size dependence of the mean polar order
 $\mean{p}$ and the mean cluster ratio $\mean{q}$ at $\rho_0=0.16$
 ($\xi = 4.43$) of the neutral swimmer $v_2=0$ \NYY{(black)}.
 Both the simulations with far- and near-field interactions and only
 near-field interactions are shown.
 \NYY{
 The results of the simulations with far- and near-field interactions
 for weak pushers $v_2=0.05$ are also shown in blue.
 }
 \NY{
 The legends are shared by all the figures.
 }
\label{fig.analysis}
}
\end{center}
\end{figure}

\tl{
Most surprisingly we find that for neutral (quadrupolar) squirmers and squirmers with small dipoles, $\left |v_2 \right | \ll 1$,  
the swimmers self-organise into a polar state with  aligned orientations and swimming directions. 
This polar order vanishes at low density.
Screening far-field interactions~\cite{Ball:1997} leads to polar order at lower densities (see Fig.~\ref{fig.analysis}(C)).
As $|v_2|$
  is increased, polar order vanishes.
 For example, for pushers ($v_2 >0$), polar order disappears at $v_2^* \simeq 0.15$
 ($\beta_{2/1} \simeq 0.58$) in Fig.~\ref{fig.analysis}(A).
 The loss of polar order is accompanied by divergence of fluctuations of the polarity
 as shown in Fig.~\ref{fig.analysis}(A).  The position of the phase boundary is not symmetric about $v_2  = 0$, i.e. different $|v_2^*|$ for  pushers  and pullers (see Fig.~\ref{fig.analysis}(B)).
 }
 
\tl{
We check the stability of polar order to fluctuations by 
 adding Gaussian white noise of amplitude $\sigma$ to the rotation in eqn. (\ref{janus.dotr.u}).
 At a fixed density, $\rho_0$ we find a transition from a gas to a polar state at a critical value of $\sigma_c>0$.
Polar order remains as we increase system size. 
In Fig.~\ref{fig.analysis}(D),  polar order is shown as a function of
the number of particles, $N$ up to $N=8192$ for the near+far field system
and $N=32768$ for the near-field-only system.
Therefore we conclude that 
the system is \tl{truly} in a state with macroscopic global polar
order.}
\NYY{
These system sizes are comparable to ABPs and the Vicsek models. 
}
\tl{The mean cluster ratio, $\aver{q}$, the fraction of swimmers in large clusters, is nearly zero throughout the polar phase (see Fig.~\ref{fig.analysis}(D)), indicating clusters are not associated with polar order.}

 

\begin{figure}[h]
\begin{center}
\includegraphics[width=0.45\textwidth]{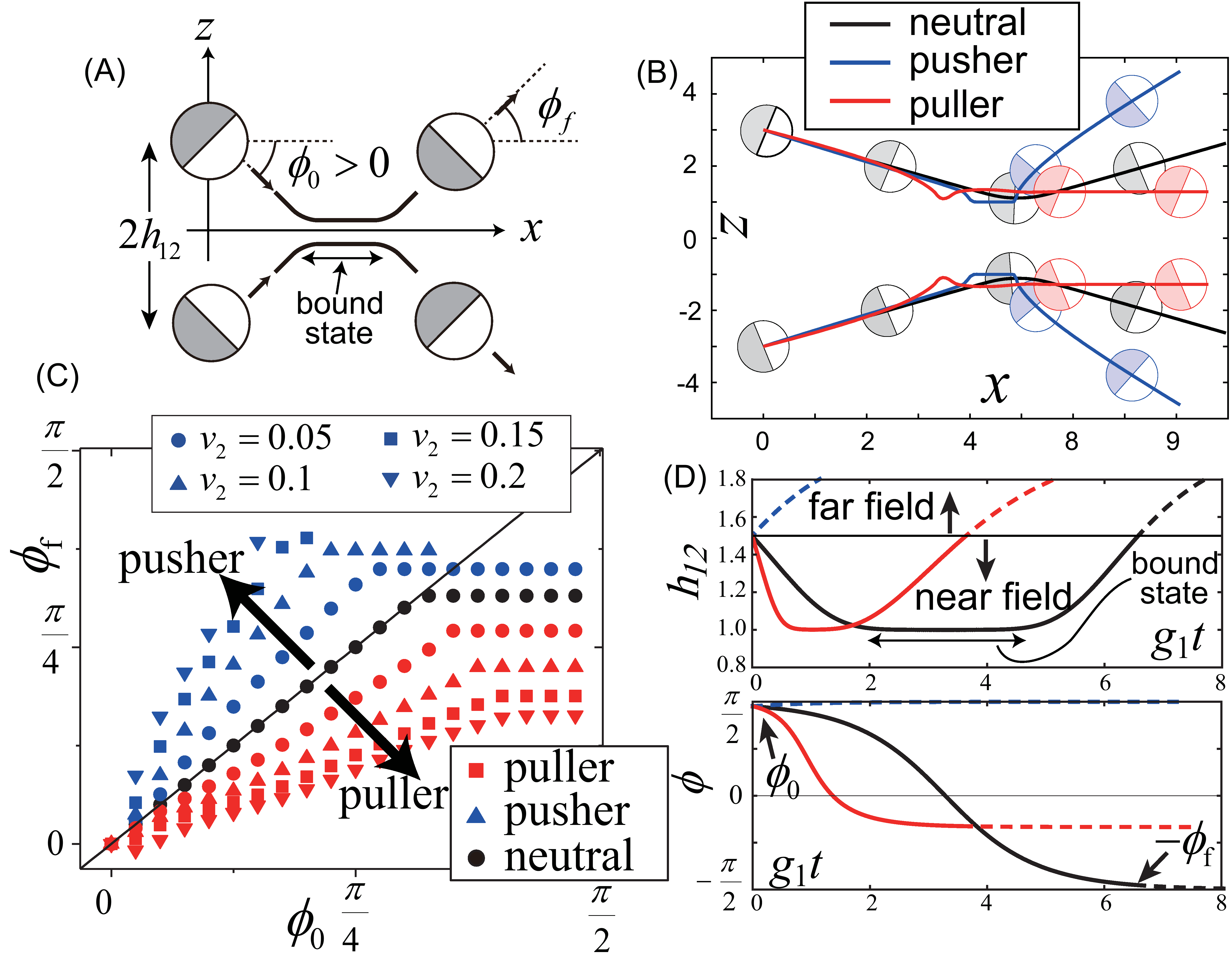}
 \caption{
 (Colour Online) (A) Schematics of two-body collisions. (B)  Typical
 trajectories  for  neutral, pusher, and puller swimmers are shown.
\NYY{
 (C) The incidence $\phi_0$ and reflection $\phi_f$ angles
 for symmetric collisions for neutral swimmers ($v_2=0$) and small
 deviation of $v_2$ to pushers ($v_2>0$) and pullers ($v_2<0$).
 The solid line shows $\phi_0 = \phi_f$.
 Only the range of $0 \leq \phi_0, \phi_f \leq \pi/2$ is shown.
 }
 (D) The dynamics of the separation $h_{12}(t)$ and the angle $\phi(t)$ in the near-field region
 for $g_2/g_1=1$.
 \NY{
 Here, $g_1 \sim u_0/R$ and $g_2\sim v_2/R$\cite{SM}.
 The time scale is normalised by $g_1^{-1}$. 
 }
 Motion outside the near-field region is indicated by dashed lines.
  \NY{
 The legends are shared by all the figures.
 }
\label{fig.two.body}
}
\end{center}
\end{figure}

It is evident from the simulations that collisions between the particles are
key in the development of polar order.
Hence, 
\NY{
we explore a two body collision in detail (see
Fig.~\ref{fig.two.body}(A) and \cite{SM}).
}
Figure \ref{fig.two.body}(B) shows some trajectories of
two `colliding' squirmers. 

\tl{
For the far-field only system, any transient alignment of pairs of squirmers is unstable to rotational fluctuations arising from 
collisions with other particles and no polar order is developed.
Including the near-field (lubrication) interaction however leads to
reorientation
\NY{
while in transient bound (Fig.~\ref{fig.two.body}(A))
states occurring during collisions as shown in
Fig.~\ref{fig.two.body}(D).
}
For small incident angles ($|\phi_0| \lesssim \pi/4$), collisions are symmetric, i.e the reflection angle ($\phi_f$)  equals $\phi_0$ but for as $|\phi_0|$ increases,  $|\phi_f|$ stops increasing and tends to a finite ({\em saturation}) angle $\lesssim \pi/4$. This asymmetry between incident and reflection angles means
($\mean{|\phi_f|} <  \mean{|\phi_0|}$, see Fig.~\ref{fig.two.body}(C)) and leads eventually to alignment.
This effect is most pronounced for neutral swimmers ($v_2=0$); while
similar behaviour is seen for  pushers and pullers,
\NYY{
 shorter residence times for pullers (Fig.~\ref{fig.two.body}(D)) and
 larger reflection angles for pushers (Fig.~\ref{fig.two.body}(C)) eventually
 destroy the polar state for both of them as $|v_2|$ becomes large.
}
 The saturation angle 
in Fig.~\ref{fig.two.body}(C) is due to 
direct contacts between squirmers (repulsive forces from the interaction potential).
This reorientation depends weakly on the contact
interaction; big changes of interaction potential lead only to slight shifts of
saturation angles. Hence the collective behaviour and phase boundaries are independent of the choice of potential~\cite{SM}. }



\tl{While we only  considered pairwise interactions, combining many of them
results in many-body effects which become relevant for a non-dilute
suspension.
In fact, for a dense suspension, 
dynamics is dominated by the lubrication interaction between swimmers which is well 
approximated by a sum of two-body interactions. 
\NYY{
In particular, $\mean{p}$ is independent of $N$ even for weak pushers
$v_2 \neq 0$ as shown in Fig.~\ref{fig.analysis}(D).
}
To understand how these many-body effects give rise to collective behaviour, we have carried out
 numerical simulations of a minimal model, in which the only non-zero interactions are  rotational near-field: $\omega_{lm}^{(ij)} \sim \epsilon^0$ ~\cite{SM} plus noise.  It has two key ingredients : short-range orientational interactions and short-range repulsive interactions.
 We are able to reproduce the same polar-disorder phase transition by increasing
 the noise amplitude. We conclude that the detailed form of the hydrodynamic interactions  are not essential for the  development of polar order. 
In contrast to the Vicsek model, here the lack of an alignment rule means 
excluded volume interactions are required to generate polar order.
 } 


\tl{The existence of polar order is fundamentally surprising because of the apparent contradiction with the well accepted generic instability of polar/nematic order of wet active matter~\cite{Simha:2002,Marchetti:2013}. 
To understand this we construct a two-fluid model for the system:  the suspending fluid (volume fraction, $1-\phi$) with velocity  $\bfv (\bfr,t)$ and the active particle (squirmer) 'phase' (volume fraction $\phi$) with local displacement variable $\bfu (\bfr,t)$ due to squirmer density variations. 
 Finally we identify a  {\em local} polar order parameter,  $\bfp (\bfr,t)$. Our analysis highlights collisions of the swimmers as essential for the formation of polar order. 
An isolated squirmer swims with velocity $u_0 \bfp$ relative to the background fluid.
The  fluid obeys the Stokes equation with a force density  $\bff^c \sim O(\rho_a^2)$ 
due to collisions between squirmers  and 
%
an active stress 
 $  
 \sigma^a_{ij} = {\zeta p_i p_j} \; 
 $
 where $\zeta \propto v_2$ and $\rho_a = \phi
\rho$  is the average density of active particles.
Replacing $\bff^c$ by $\chi \left( \dot \bfu - \bfv - u_0 \bfp
\right)$. 
and linearising $\bfp,\bfv$ about the homogeneous state: $\bfv_0 = \dot\bfu_0 - u_0 \bfp_0 $, 
\begin{eqnarray}
0 &=& \eta \nabla^2 \delta \bfv + \chi \delta \bfv - \nabla P + \nabla \cdot {\bsigma}^a \; ; \; \nabla \cdot  \delta \bfv = 0 \;,  \\
 \partial_t \delta \bfp &=& - u_0 \bfp_0 \cdot \nabla \delta \bfp +
 \delta \bomega \cdot  \bfp_0 + \gamma \delta \bfe \cdot \bfp_0 + K \nabla^2 \delta \bfp  \quad 
\end{eqnarray}
where $\omega_{ij} = \frac12 ( \partial_i v_j - \partial_j v_i)$, $e_{ij} = \frac12 ( \partial_i v_j + \partial_j v_i)$ and $K$ is the Frank elastic constant in the one constant approximation.
A finite screening length $\xi \sim \sqrt{\eta/\chi}$ weakens the generic instability from long-wavelengths to finite wavelengths and stabilises the polar state on long lengthscales. Hence a comparison between the screening length $\xi$ and the active lengthscale, $\sqrt{K/|\zeta|}$~\cite{Voituriez:2005}, allows us to determine the onset of polar order for $|\zeta| < \zeta_c = K / \xi^2$, i.e. for swimmers that are close to neutral. We recover the generic instability as $\phi \rar 0$.
}



\tl{
We emphasise that the computational expense of including hydrodynamics in the simulations of collective behaviour of active matter requires trade-offs where accuracy is sacrificed. Navier-Stokes (NS) solvers such as Lattice-Boltzmann compromise on the resolution of the velocity field and hence do not accurately describe fluid flow when the active particles are very close to each other. Here we have developed another scheme whose strengths are exactly where NS solvers are weak, for active particles in close proximity. It is also very accurate when the particles are well separated. Where it is less accurate, however is at intermediate separations. Its other great advantage is the ease with which we can study  
systems  with many more particles. 
Another nice feature is the ability to switch off different contributions to the motion to identify the dominant mechanisms behind the macroscopic phenomena observed. Using it we have studied the collective behaviour of large numbers of spherical active particles and confirmed and clarified the phenomena observed in smaller simulations. Dense cluster formation is suppressed and we show that it is replaced by open gel-like aggregates at higher densities and most surprisingly, a polar ordered phase is stabilised by hydrodynamic lubrication interactions. We have also provided  analytic continuum arguments explaining how such a state can be realised.
 In addition to the  work presented here, we have studied 
 purely 2D systems (2D with 2D interactions) and 3D systems (3D
with 3D interactions), and obtained similar results, but at higher densities\cite{inpreparation}.
}


\begin{acknowledgments}
The authors are  grateful to S. Fielding, T. Ishikawa and R. Golestanian for
 helpful discussions.
  NY acknowledges the support by JSPS KAKENHI grant numbers JP26800219,
 JP26103503, and JP16H00793.
 NY also acknowledges the support by JSPS A3 Foresight Program.
 TBL is supported by BrisSynBio, a BBSRC/EPSRC Advanced Synthetic Biology
 Research Center (grant number BB/L01386X/1).
We would like to thank the Isaac Newton Institute for Mathematical Sciences, Cambridge, for support and hospitality during the programmes, ``The Mathematics of Liquid Crystals'' and ``Dynamics of active suspensions, gels, cells and tissues'' where work on this article was started.
\end{acknowledgments}



\begin{thebibliography}{10}

\bibitem{Marchetti:2013}
M.~C. Marchetti, J.~F. Joanny, S.~Ramaswamy, T.~B. Liverpool, J.~Prost, Madan
  Rao, and R.~Aditi Simha.
\newblock Hydrodynamics of soft active matter.
\newblock {\em Rev. Mod. Phys.}, 85:1143--1189, 2013.

\bibitem{Toner2005}
J.~Toner, Y.~Tu, and S.~Ramaswamy.
\newblock Hydrodynamics and phases of flocks.
\newblock {\em Ann. Phys.}, 318(1):170--244, 2005.

\bibitem{ramaswamy:2010}
Sriram Ramaswamy.
\newblock The mechanics and statistics of active matter.
\newblock {\em Ann. Rev. Cond. Mat. Phys.}, 1:323--345, 2010.

\bibitem{vicsek:1995}
Tam\'as Vicsek, Andr\'as Czir\'ok, Eshel Ben-Jacob, Inon Cohen, and Ofer
  Shochet.
\newblock Novel type of phase transition in a system of self-driven particles.
\newblock {\em Phys. Rev. Lett.}, 75(6):1226--1229, Aug 1995.

\bibitem{Henricus:2012}
Henricus~H. Wensink, J{\"{o}}rn Dunkel, Sebastian Heidenreich, Knut Drescher,
  Raymond~E. Goldstein, Hartmut L{\"{o}}wen, and Julia~M. Yeomans.
\newblock Meso-scale turbulence in living fluids.
\newblock {\em Proc. Nat. Acad. Sci.}, 109(36):14308--14313, 2012.

\bibitem{Cates:2012}
M~E Cates.
\newblock Diffusive transport without detailed balance in motile bacteria: does
  microbiology need statistical physics?
\newblock {\em Rep. Prog. Phys.}, 75(4):042601, 2012.

\bibitem{Palacci:2013}
Jeremie Palacci, Stefano Sacanna, Asher~Preska Steinberg, David~J. Pine, and
  Paul~M. Chaikin.
\newblock Living crystals of light-activated colloidal surfers.
\newblock {\em Science}, 339(6122):936--940, 2013.

\bibitem{Bricard:2013}
Antoine Bricard, Jean-Baptiste Caussin, Nicolas Desreumaux, Olivier Dauchot,
  and Denis Bartolo.
\newblock Emergence of macroscopic directed motion in populations of motile
  colloids.
\newblock {\em Nature}, 503(7474):95--98, 2013.

\bibitem{Bertin:2013a}
Eric Bertin, Hugues Chat{\'{e}}, Francesco Ginelli, Shradha Mishra, Anton
  Peshkov, and Sriram Ramaswamy.
\newblock Mesoscopic theory for fluctuating active nematics.
\newblock {\em New Journal of Physics}, 15(8):085032, 2013.

\bibitem{Fily:2012}
Yaouen Fily and M.~Cristina Marchetti.
\newblock Athermal phase separation of self-propelled particles with no
  alignment.
\newblock {\em Phys. Rev. Lett.}, 108:235702, Jun 2012.

\bibitem{Buttinoni:2013}
Ivo Buttinoni, Julian Bialk\'e, Felix K\"ummel, Hartmut L\"owen, Clemens
  Bechinger, and Thomas Speck.
\newblock Dynamical clustering and phase separation in suspensions of
  self-propelled colloidal particles.
\newblock {\em Phys. Rev. Lett.}, 110:238301, Jun 2013.

\bibitem{Redner:2013}
Gabriel~S. Redner, Michael~F. Hagan, and Aparna Baskaran.
\newblock Structure and dynamics of a phase-separating active colloidal fluid.
\newblock {\em Phys. Rev. Lett.}, 110:055701, 2013.

\bibitem{Lighthill:1952}
M.~J. Lighthill.
\newblock On the squirming motion of nearly spherical deformable bodies through
  liquids at very small reynolds numbers.
\newblock {\em Communications on Pure and Applied Mathematics}, 5(2):109--118,
  1952.

\bibitem{blake:1971}
JR~Blake.
\newblock Self propulsion due to oscillations on the surface of a cylinder at
  low reynolds number.
\newblock {\em Bulletin of the Australian Mathematical Society},
  5(02):255--264, 1971.

\bibitem{Pak:2014}
OnShun Pak and Eric Lauga.
\newblock Generalized squirming motion of a sphere.
\newblock {\em Journal of Engineering Mathematics}, 88(1):1--28, 2014.

\bibitem{golestanian:2005}
Ramin Golestanian, Tanniemola~B. Liverpool, and Armand Ajdari.
\newblock Propulsion of a molecular machine by asymmetric distribution of
  reaction products.
\newblock {\em Physical Review Letters}, 94(22):220801, 2005.

\bibitem{ishikawa:2006}
T.~Ishikawa, MP~Simmonds, and TJ~Pedley.
\newblock Hydrodynamic interaction of two swimming model micro-organisms.
\newblock {\em J. Fluid Mech.}, 568:119--160, 2006.

\bibitem{Llopis:2006}
I.~Llopis and I.~Pagonabarraga.
\newblock Dynamic regimes of hydrodynamically coupled self-propelling
  particles.
\newblock {\em Eur. Phys. Lett.}, 75(6):999, 2006.

\bibitem{Ishikawa:2008a}
Takuji Ishikawa, J.~T. Locsei, and T.~J. Pedley.
\newblock Development of coherent structures in concentrated suspensions of
  swimming model micro-organisms.
\newblock {\em J. Fluid Mech.}, 615(-1):401--431, 2008.

\bibitem{Ishimoto:2013}
Kenta Ishimoto and Eamonn~A. Gaffney.
\newblock Squirmer dynamics near a boundary.
\newblock {\em Phys. Rev. E}, 88:062702, Dec 2013.

\bibitem{Li:2014}
Gao-Jin Li and Arezoo~M. Ardekani.
\newblock Hydrodynamic interaction of microswimmers near a wall.
\newblock {\em Phys. Rev. E}, 90:013010, Jul 2014.

\bibitem{SharifiMood:2015}
Nima Sharifi-Mood, Ali Mozaffari, and Ubaldo C{\'{o}}rdova-Figueroa.
\newblock Pair interaction of catalytically active colloids: From assembly to
  escape.
\newblock {\em arXiv:1510.03000}, 2015.

\bibitem{Papavassiliou:2016}
Dario Papavassiliou and Gareth~P. Alexander.
\newblock Exact solutions for hydrodynamic interactions of two squirming
  spheres.
\newblock {\em arXiv:1602.06912}, 2016.

\bibitem{Mucha:2004}
Peter~J. Mucha, Shang-You Tee, David~A. Weitz, Boris~I. Shraiman, and
  Michael~P. Brenner.
\newblock A model for velocity fluctuations in sedimentation.
\newblock {\em Journal of Fluid Mechanics}, 501:71--104, 02 2004.

\bibitem{Molina:2013}
John~J. Molina, Yasuya Nakayama, and Ryoichi Yamamoto.
\newblock Hydrodynamic interactions of self-propelled swimmers.
\newblock {\em Soft Matter}, 9:4923--4936, 2013.

\bibitem{Zoettl:2014}
Andreas Z{\"{o}}ttl and Holger Stark.
\newblock Hydrodynamics determines collective motion and phase behavior of
  active colloids in quasi-two-dimensional confinement.
\newblock {\em Phys. Rev. Lett.}, 112:118101, Mar 2014.

\bibitem{Matas-Navarro:2014}
Ricard Matas-Navarro, Ramin Golestanian, Tanniemola~B. Liverpool, and
  Suzanne~M. Fielding.
\newblock Hydrodynamic suppression of phase separation in active suspensions.
\newblock {\em Phys. Rev. E}, 90:032304, Sep 2014.

\bibitem{Delfau:2016}
J.-B. Delfau, J.~Molina, and M.~Sano.
\newblock Collective behavior of strongly confined suspensions of squirmers.
\newblock {\em Eur. Phys. Lett.}, 114(2):24001, 2016.

\bibitem{Spagnolie:2012}
Saverio~E Spagnolie and Eric Lauga.
\newblock Hydrodynamics of self-propulsion near a boundary: predictions and
  accuracy of far-field approximations.
\newblock {\em Journal of Fluid Mechanics}, 700:105--147, 2012.

\bibitem{Saintillan:2015}
D.~Saintillan and M.~J. Shelley.
\newblock {\em Complex Fluids in Biological Systems}, chapter Theory of active
  suspensions, pages 319--355.
\newblock Springer, 2015.

\bibitem{Theurkauff:2012}
I.~Theurkauff, C.~Cottin-Bizonne, J.~Palacci, C.~Ybert, and L.~Bocquet.
\newblock Dynamic clustering in active colloidal suspensions with chemical
  signaling.
\newblock {\em Phys. Rev. Lett.}, 108:268303, Jun 2012.

\bibitem{peruani:2006}
Fernando Peruani, Andreas Deutsch, and Markus B\"ar.
\newblock Nonequilibrium clustering of self-propelled rods.
\newblock {\em Phys. Rev. E}, 74(3):030904, Sep 2006.

\bibitem{Blaschke:2016}
Johannes Blaschke, Maurice Maurer, Karthik Menon, Andreas Z{\"{o}}ttl, and
  Holger Stark.
\newblock Phase separation and coexistence of hydrodynamically interacting
  microswimmers.
\newblock {\em Soft Matter}, 12:9821--9831, 2016.

\bibitem{Saffman:1975}
P~G Saffman and M~Delbrück.
\newblock Brownian motion in biological membranes.
\newblock {\em Proc. Nat. Acad. Sci.}, 72(8):3111--3113, 1975.

\bibitem{Saha:2013}
Suropriya Saha, Ramin Golestanian, and Sriram Ramaswamy.
\newblock Clusters, asters, and collective oscillations in chemotactic
  colloids.
\newblock {\em Phys. Rev. E}, 89:062316, Jun 2014.

\bibitem{Navarro:2015}
Ricard~Matas Navarro and Suzanne~M. Fielding.
\newblock Clustering and phase behaviour of attractive active particles with
  hydrodynamics.
\newblock {\em Soft Matter}, 11:7525--7546, 2015.

\bibitem{Alarcon:2016}
Francisco Alarcon, Chantal Valeriani, and Ignacio Pagonabarraga.
\newblock Morphology of clusters of attractive dry and wet self-propelled
  spherical particle suspensions.
\newblock {\em Soft Matter}, pages~--, 2016.

\bibitem{Evans:2011}
Arthur~A. Evans, Takuji Ishikawa, Takami Yamaguchi, and Eric Lauga.
\newblock Orientational order in concentrated suspensions of spherical
  microswimmers.
\newblock {\em Physics of Fluids}, 23(11):--, 2011.

\bibitem{Alarcon:2013}
F.~Alarc{\'{o}}n and I.~Pagonabarraga.
\newblock Spontaneous aggregation and global polar ordering in squirmer
  suspensions.
\newblock {\em Journal of Molecular Liquids}, 185:56 -- 61, 2013.
\newblock Molecular Simulations of Complex Systems.

\bibitem{Oyama:2016}
John Jairo~Molina Norihiro~Oyama and Ryoichi Yamamoto.
\newblock A binary collision route for purely hydrodynamic orientational
  ordering of microswimmers.
\newblock {\em arXiv:1606.03839}, 2016.

\bibitem{Swan:2011}
James~W. Swan, John~F. Brady, Rachel~S. Moore, and ChE 174.
\newblock Modeling hydrodynamic self-propulsion with stokesian dynamics. or
  teaching stokesian dynamics to swim.
\newblock {\em Physics of Fluids}, 23(7), 2011.

\bibitem{Hill:1954}
E.~L. Hill.
\newblock The theory of vector spherical harmonics.
\newblock {\em Am. J. Phys.}, 22(4):211--214, 1954.

\bibitem{SM}
See Supplemental Material at
  http://link.aps.org/supplemental/???.?????? for details
  of the fomula and the numerical simulations.

\bibitem{Jeffrey:1984}
DJ~Jeffrey and Y~Onishi.
\newblock Calculation of the resistance and mobility functions for two unequal
  rigid spheres in low-reynolds-number flow.
\newblock {\em Journal of Fluid Mechanics}, 139:261--290, 1984.

\bibitem{Kim:1991}
S.~Kim and S.J. Karrila.
\newblock {\em Microhydrodynamics}.
\newblock Butterworth-Heinemann New York, 1991.

\bibitem{Ball:1997}
R.C. Ball and J.R. Melrose.
\newblock A simulation technique for many spheres in quasi-static motion under
  frame-invariant pair drag and brownian forces.
\newblock {\em Physica A}, 247(1-4):444 -- 472, 1997.

\bibitem{Simha:2002}
R.~Aditi Simha and Sriram Ramaswamy.
\newblock Hydrodynamic fluctuations and instabilities in ordered suspensions of
  self-propelled particles.
\newblock {\em Phys. Rev. Lett.}, 89(5):058101, Jul 2002.

\bibitem{Voituriez:2005}
R~Voituriez, JF~Joanny, and J~Prost.
\newblock Spontaneous flow transition in active polar gels.
\newblock {\em EPL (Europhysics Letters)}, 70(3):404, 2005.

\bibitem{inpreparation}
N. Yoshinaga and T. B. Liverpool (unpublished).

\end{thebibliography}

\pagebreak

\appendix

\end{document}